%Paper: 9203021
%From: Hiroshi Suzuki <SUZUKI%JPNRIFP.BITNET@pucc.princeton.edu>
%Date: Mon, 09 Mar 92 15:03:43 JST

\input phyzzx
%%%%%%%%%%%%%%%%%%%%%%%%%%%%%%%%%%%%%%%%%%%%%%%%%%%%%%%%%%%%%%%%%
%
% The following is a PHYZZX file, so please compile by a PHYZZX
% preloaded TeX or please insert, input PHYZZX in the top of the file.
%
%%%%%%%%%%%%%%%%%%%%%%%%%%%%%%%%%%%%%%%%%%%%%%%%%%%%%%%%%%%%%%%%%
\def\Dslash{\,\raise.15ex\hbox{/}\mkern-13.5mu D}
\def\ee{\eqno\eq }
%%%%%%%%%%%%%%%%%%%%%%%%%%%%%%%%%%%%%%%%%%%%%%%%%%%%%%%%%%%%%%%%%%%%
\REF\ADE{
 M.\ Ademollo, L.\ Brink, A.\ D'Adda, R.\ D'Auria, E.\ Napolitano,
 S.\ Sciuto, E.\ Del Giudice, P.\ Di Vecchia, S.\ Ferrara, F.\ Gliozzi,
 R.\ Musto and R.\ Pettorino,
    {\sl Phys.\ Lett.} {\bf B62} (1976) 105;\hfil\break
 M.\ Ademollo, L.\ Brink, A.\ D'Adda, R.\ D'Auria, E.\ Napolitano,
 S.\ Sciuto, E.\ Del Giudice, P.\ Di Vecchia, S.\ Ferrara, F.\ Gliozzi,
 R.\ Musto, R.\ Pettorino and J.\ H. Schwarz,
    {\sl Nucl.\ Phys.} {\bf B111} (1976) 77.}
\REF\BRU{
 D.\ J.\ Bruth, D.\ B.\ Fairle and R.\ G.\ Yates,
    {\sl Nucl.\ Phys.} {\bf 108} (1976) 310;\hfil\break
 L.\ Brink and J.\ H.\ Schwarz,
    {\sl Nucl.\ Phys.} {\bf B121} (1977) 285;\hfil\break
 A.\ D'Adda and F.\ Lizzi,
    {\sl Phys.\ Lett.} {\bf B191} (1987) 85;\hfil\break
 S.\ D.\ Mathur ans S.\ Mukhi,
    {\sl Nucl.\ Phys.} {\bf B302} (1988) 130;\hfil\break
 N.\ Ohta and S.\ Osabe,
    {\sl Phys.\ Rev.} {\bf D39} (1989) 1641;\hfil\break
 M.\ Corvi, V.\ A.\ Kosteleck\'y and P.\ Moxhay,
    {\sl Phys.\ Rev.} {\bf D39} (1989) 1611.}
\REF\FRAD{
 E.\ S.\ Fradkin and A.\ A.\ Tseytlin,
    {\sl Phys.\ Lett.} {\bf B106} (1981) 63.}
\REF\BOUW{
 P.\ Bouwknegt and P.\ van Nieuwenhuizen,
    {\sl Class.\ Quantum Grav.} {\bf 3} (1986) 207.}
\REF\BIL{
 A.\ Bilal,
    {\sl Phys.\ Lett.} {\bf B180} (1986) 255;\hfil\break
 A.\ R.\ Bogojevic and Z.\ Hlousek,
    {\sl Phys.\ Lett.} {\bf B179} (1986) 69;\hfil\break
 S.\ D.\ Mathur ans S.\ Mukhi,
    {\sl Phys.\ Rev.} {\bf D36} (1987) 465.}
\REF\GOM{
 J.\ Gomis,
    {\sl Phys.\ Rev.} {\bf D40} (1989) 408.}
\REF\OOG{
 H.\ Ooguri and C.\ Vafa,
    {\sl Nucl.\ Phys.} {\bf B361} (1991) 469.}
\REF\WHO{
 J.\ Bie\'nkowska, Chicago preprint, EFI 91-65.}
\REF\DIS{
 J.\ Distler, Z.\ Hlousek and H.\ Kawai,
    {\sl Int.\ J.\ Mod.\ Phys.} {\bf A5} (1990) 391;\hfil\break
 I.\ Antoniadis, C.\ Bachas and C.\ Kounnas,
    {\sl Phys.\ Lett.} {\bf B242} (1990) 185.}
\REF\WITT{
 E.\ Witten,
    {\sl Commun.\ Math.\ Phys.} {\bf 117} (1988) 353.}
\REF\WIT{
 E.\ Witten,
    {\sl Commun.\ Math.\ Phys.} {\bf 118} (1988) 411; {\sl Nucl.
\ Phys.} {\bf B340} (1990) 281.}
\REF\OUR{
 J.\ Gomis and H.\ Suzuki, YITP preprint, YITP/U-91-53, to appear in
 {\sl
 Phys.\ Lett.\ B}.}
\REF\EGU{
 T.~Eguchi and S.-K.~Yang,
    {\sl Mod.~Phys.~Lett.} {\bf A5} (1990) 1693.}
\REF\ANOMALY{
 K. Fujikawa, {\sl Phys.\ Rev.\ Lett.} {\bf 42} (1979) 1195; {\bf 44}
 (1980) 1733; {\sl Phys.\ Rev.\/} {\bf D21} (1980) 2848.}
\REF\ROCV{
 M. Roc\v ek, P. van Nieuwenhuizen and S. C. Zhang, {\sl Ann.\ Phys.}
 (NY) {\bf 172} (1986) 348.}
\REF\YOUR{
 K. Fujikawa, T. Inagaki and H. Suzuki,
    {\sl Nucl.\ Phys.} {\bf B332} (1990) 499.}
\REF\YASUDA{
 K. Fujikawa, {\sl Phys.\ Rev.} {\bf D25} (1982) 2584; {\sl Nucl.
\ Phys.} {\bf B226} (1983) 437;\hfil\break
 K. Fujikawa and O. Yasuda, {\sl Nucl.\ Phys.} {\bf B245} (1984)
 436.}
\REF\GRAVITYTWO{
 K. Fujikawa, U. Linstr\"om, N. K. Nielsen, M. Roc\v ek and P. van
 Nieuwenhuizen, {\sl Phys.\ Rev.} {\bf D37} (1988) 391.}
\REF\FUJ{
 K.\ Fujikawa,
    {\sl Phys.\ Rev.} {\bf D25} (1982) 2584.}
\REF\FRI{
 D.\ Friedan, E.\ Martinec and S.\ Shenker,
    {\sl Nucl.\ Phys.} {\bf B271} (1986) 93.}
\REF\YOU{
 K.\ Fujikawa, T.\ Inagaki and H.\ Suzuki,
    {\sl Phys.\ Lett.} {\bf B213} (1988) 279.}
\REF\NOJ{
 S.\ Nojiri,
    {\sl Phys.\ Lett.} {\bf B262} (1991) 419; {\sl Phys.\ Lett.} {\bf
 B264} (1991) 57.}
\REF\FUJI{
 K.\ Fujikawa and H.\ Suzuki,
    {\sl Nucl.\ Phys.} {\bf B361} (1991) 539.}
%%%%%%%%%%%%%%%%%%%%%%%%%%%%%%%%%%%%%%%%%%%%%%%%%%%%%%%%%%%%%%%%%%%%%%
%\REF\MAR{
% S.\ P.\ Martin,
%    {\sl Phys.\ Lett.} {\bf B191} (1987) 81.}
%\REF\LI{
% K.\ Li,
%    {\sl Nucl.\ Phys.} {\bf B346} (1990) 329.}
%\REF\ROC{
% J.\ Gomis and J.\ Roca,
%    {\sl Phys.\ Lett.} {\bf B268} (1991) 197.}
%\REF\NEL{
% S.\ Govindarajan, P.\ Nelson and S.-J.\ Rey,
%    {\sl Nucl.\ Phys.} {\bf B365} (1991) 633.}
%%%%%%%%%%%%%%%%%%%%%%%%%%%%%%%%%
%%%%%%%%%%%%%%%%%%%%%%%%%%%%%%%%%%%%%%%%%%%%%%%%%%
\def\partialbar{\overline\partial}
\def\Xmu{X^\mu}
\def\Ymu{Y^\mu}
\def\psiplusmu{\psi^{+\mu}}
\def\psiminusmu{\psi^{-\mu}}
\def\betaplus{\beta^+}
\def\betaminus{\beta^-}
\def\gammaplus{\gamma^+}
\def\gammaminus{\gamma^-}
\def\Di{{\cal D}}
\def\bl{\bigl(}
\def\br{\bigr)}
\def\thetaplus{\theta^+}
\def\thetaminus{\theta^-}
\def\Diplus{D^+}
\def\Diminus{D^-}
%%%%%%%%%%%%%%%%%%%%%%%%%%%%%%%%%%%%%%%%%%%%%%%%%%%%%%%%%%%%%%%%%%%%
\pubnum={92-4\cr
UB-ECM-PF 92/4}
\date={February 1992}

\titlepage
\title{Covariant Currents in $N=2$ Super-Liouville Theory}

\author{Joaquim Gomis\footnote{\ast}{Electronic address:
 quim@ebubecm1.bitnet}%
and Hiroshi Suzuki\footnote{\dagger}{JSPS Junior
 Scientist Fellow. Also at Department of Physics, Hiroshima University,
 Higashi-Hiroshima 724, Japan. Electronic address:
 suzuki@jpnrifp.bitnet}}

\address{$^\ast$Department d'Estructura i Constituents de la
 Mat\`eria\break
Universitat de Barcelona, Diagonal 647, E-08028 Barcelona, Catalonia,
 Spain}

\address{$^\dagger$Uji Research Center\break
Yukawa Institute for Theoretical Physics\break
Kyoto University, Uji 611, Japan}

\abstract{Based on a path integral prescription for anomaly
 calculation, we analyze an effective theory of the two-dimensional
 $N=2$ supergravity, i.e., $N=2$ super-Liouville theory. We calculate
 the anomalies associated with the BRST supercurrent and the ghost
 number supercurrent. From those expressions of anomalies, we construct
 covariant BRST and ghost number supercurrents in the effective theory.
 We then show that the (super-)coordinate BRST current algebra forms a
 superfield extension of the topological conformal algebra for an {\it
 arbitrary\/} type of conformal matter or, in terms of the string
 theory, for an arbitrary number of space-time dimensions. This fact is
 very contrast with $N=0$ and $N=1$ (super-)Liouville theory, where the
 topological algebra singles out a particular value of dimensions. Our
 observation suggests a topological nature of the two-dimensional $N=2$
 supergravity as a quantum theory.}
%%%%%%%%%%%%%%%%%%%%%%%%%%%%%%%%%
%%%%%%%   CONTENTS Starts %%%%%%%
%%%%%%%%%%%%%%%%%%%%%%%%%%%%%%%%%
\chapter{Introduction}

The $N=2$ string or the two-dimensional $N=2$ supergravity introduced
 by Ademollo {\it et al\/} [\ADE,\BRU,\FRAD,\BOUW,\BIL,\GOM,\OOG] has
 critical dimension $d=2$ and there is no transverse degree of freedom.
 Very recently, it has been argued that the no-ghost theorem can be
 established [\WHO]. The $N=2$ subcritical strings or $N=2$
 super-Liouville theory has also been analyzed. Distler, Hlousek, and
 Kawai [\DIS] noticed that the local ansatz for the Jacobian, that
 relates the interacting measure with the free measure, works for any
 kind of conformal matter or, in terms of strings, for an arbitrary
 number of space-time dimensions. All those special features of $N=2$
 string suggest that this theory is a topological quantum field theory
 [\WITT,\WIT]. In a previous work [\OUR], we proved that critical and
 subcritical $N=2$ strings are topological field theories in the sense
 that the (super-)coordinate BRST current algebra gives a realization
 of an $N=2$ superfield extension of the topological conformal algebra
 [\WITT,\EGU] for arbitrary type of conformal matter.

In this paper we want to analyze in detail the appearance of this
 topological conformal algebra in the case of $N=2$ super-Liouville
 theory by constructing an effective theory based on anomalous
 identities  associated with the BRST and ghost number symmetries. The
 finite renormalization of the coupling constant will be interpreted as
 a one-loop order effect of the BRST invariant measure. The relation
 with the critical string will be also commented.

The organization of the paper is as follows. In Section 2, we compute
 the BRST and ghost number anomalies in the $N=2$ supergravity in the
 superconformal gauge, by using the path integral representation
 [\ANOMALY]. The $N=2$ super-Liouville theory is constructed in Section
 3 with a consideration on the covariant BRST and ghost number
 supercurrents. In Section 4 we derive the topological conformal
 algebra. In section 5 there are some conclusions and there is an
 Appendix about some basic facts of the $N=2$ superfield formalism.

\chapter{BRST and ghost number anomalies in $N=2$ supergravity}

In this section, we compute anomalies associated with the BRST and the
 ghost number supercurrents in the two-dimensional $N=2$ supergravity.
 We follow the path integral prescription for anomaly calculation
 [\ANOMALY]. In the case of $N=1$ supergravity, the superfield path
 integral is known to be an appropriate tool to compute those anomalies
 [\ROCV,\YOUR]. For $N=2$ case, however, it is well-known that the
 action of the matter multiplet (A.4) cannot be written directly in
 terms of the real scalar superfield (A.8) [\ADE]. Therefore it is not
 clear how the superfield path integral is useful in the present
 context of the anomaly calculation\rlap.\footnote{\star}{For the ghost
 and anti-ghost multiplet, the action can be written directly by the
 superfield as in (A.12). Thus the superfield path integral may work.}

Here we consider the path integral in the component fields. The
 classical gauge symmetries of the $N=2$ supergravity are the general
 coordinate, the local Lorentz, the $N=2$ supersymmetry, the Weyl, the
 super-Weyl, and the chiral transformations [\FRAD]. We assume under
 our regularization, the Weyl, the super-Weyl and the chiral
 transformations are anomalous at quantum level. Using the
 non-anomalous transformations, we can fix the $N=2$ supergravity
 multiplet in the superconformal gauge as
$$
\eqalign{
  &e_\mu^a=e^{\sigma/2}\delta^a_\mu,
\cr
  &\chi^\pm_\mu=\gamma_\mu\phi^\pm,
\cr
  &A_\mu=\epsilon_{\mu\nu}\partial^\nu\phi,
\cr
}
\ee
$$
where $\sigma$ is the Liouville mode and, $\phi^\pm$ and $\phi$ are
 their $N=2$ superpartners.

The BRST supercurrent and ghost number supercurrent anomalies will
 depend on the Liouville mode $\sigma$ and their $N=2$ superpartners.
 As we are working in a path integral in the components, we should
 construct the integration variables depending on $\sigma$, $\phi^\pm$,
 and $\phi$ such that the integration measure is invariant under the
 supercoordinate transformations. This program have some difficulties,
 for $N=1$, see for example [\ROCV]. Here instead we are going to use
 the following strategy. Let us forget about the anomalous character of
 the super-Weyl and the chiral transformations. By using these
 symmetries,
 we can fix the gauge
$$
\eqalign{
  &e_\mu^a=e^{\sigma/2}\delta^a_\mu,
\cr
  &\chi^\pm_\mu=0,
\cr
  &A_\mu=0.
\cr
}
\ee
$$
In this way, the BRST supercurrent and ghost number supercurrent
 anomalies will depend only on the Liouville mode. When we consider the
 effect of the super-Weyl and the chiral anomalies, the remaining
 components of the gravitino and the U(1) gauge field should appear in
 various anomalies. Here we {\it assume\/} that we are actually using a
 regularization that is invariant under the super-coordinate and the
 gauge transformations. Especially, we assume the invariance under the
 {\it global\/} super transformation:
$$
\eqalign{
  &\delta\sigma=i\left(\alpha^+\phi^--\alpha^-\phi^+\right),
\cr
  &\delta\phi=\alpha^-\phi^++\alpha^+\phi^-,
\cr
  &\delta\phi^\pm
  =\alpha^\pm\left(\partial\phi\pm i\partial\sigma\right).
\cr
}
\eqn\twotwo
$$
To get the
 dependences on the remaining components of the Liouville
 superpartners, we will use the invariance (or covariance) under
 \twotwo. This strategy is the same as the one for a calculation of the
 super-Liouville action in [\FRAD].

The partition function of the $N=2$ supergravity in the superconformal
 gauge is defined by
$$
\eqalign{
  \int\,d\widetilde\mu\,\exp\biggl\{-{1\over2\pi}\int\biggl[\,
  {1\over2}&\left(
  -\partial\Xmu\partialbar\Xmu
  -\partial\Ymu\partialbar\Ymu
  +\psiplusmu\partialbar\psiminusmu+\psiminusmu\partialbar\psiplusmu
  \right)
\cr
  &\qquad+b\partialbar c+\betaplus\partialbar\gammaminus
   +\betaminus\partialbar\gammaplus+\eta\partialbar\xi
   +({\rm c.\ c.})\,\biggr]\biggr\},
\cr
}
\eqn\twofour
$$
where we defined the integration measure $d\widetilde\mu$ by
$$
\eqalign{
  d\widetilde\mu&=
%  \Di\bl e^{\sigma/2}\br
  \Di\bl e^\sigma c\br
  \Di\bl e^\sigma\overline c\br
  \Di\bl e^{-\sigma/2}b\br
  \Di\bl e^{-\sigma/2}\overline b\br
  \Di\bl e^{3\sigma/4}\gammaplus\br
  \Di\bl e^{3\sigma/4}\overline\gammaplus\br
\cr
  &\quad\times
  \Di\bl e^{3\sigma/4}\gammaminus\br
  \Di\bl e^{3\sigma/4}\overline\gammaminus\br
  \Di\bl e^{-\sigma/4}\betaplus\br
  \Di\bl e^{-\sigma/4}\overline\betaplus\br
  \Di\bl e^{-\sigma/4}\betaminus\br
  \Di\bl e^{-\sigma/4}\overline\betaminus\br
\cr
  &\quad\times
  \Di\bl e^{\sigma/2}\xi\br
  \Di\bl e^{\sigma/2}\overline\xi\br
  \Di\eta
  \Di\overline\eta
  \Di\bl e^{\sigma/4}\psiplusmu\br
  \Di\bl e^{\sigma/4}\overline\psiplusmu\br
\cr
  &\quad\times
  \Di\bl e^{\sigma/4}\psiminusmu\br
  \Di\bl e^{\sigma/4}\overline\psiminusmu\br
  \Di\bl e^{\sigma/2}\Xmu\br
  \Di\bl e^{\sigma/2}\Ymu\br
\cr
  &\equiv
%  \Di\bl e^{\sigma/2}\br
  \Di\widetilde c\,
  \Di\widetilde{\overline c}\,
  \Di\widetilde b\,
  \Di\widetilde{\overline b}\,
  \Di\widetilde\gamma^+
  \Di\widetilde{\overline\gamma}^+
  \Di\widetilde\gamma^-
  \Di\widetilde{\overline\gamma}^-
  \Di\widetilde\beta^+
  \Di\widetilde{\overline\beta}^+
  \Di\widetilde\beta^-
  \Di\widetilde{\overline\beta}^-
  \Di\widetilde\xi
  \Di\widetilde{\overline\xi}
  \Di\eta
  \Di\overline\eta
\cr
  &\qquad\times
  \Di\widetilde\psi^{+\mu}
  \Di\widetilde{\overline\psi}^{+\mu}
  \Di\widetilde\psi^{-\mu}
  \Di\widetilde{\overline\psi}^{-\mu}
  \Di\widetilde X^\mu
  \Di\widetilde Y^\mu.
\cr
}
\eqn\twofive
$$
The various weight factors ($\exp\sigma$) in the integration measure
 are determined from the general coordinate (BRST) invariance of the
 integration measure [\YASUDA,\GRAVITYTWO]. In the above
 expression, we did not include the integration of the Liouville (or
 Weyl) mode~$\sigma$. We will turn this point in the next section. Our
 starting point \twofour\ and~\twofive\ are the same as the one in
 [\BOUW].

We also note the above integration measure is invariant under the
 conformal transformation as noted in [\YOUR]:
$$
  \delta\phi=V\partial\phi+h\left(\partial V\right)\phi,\quad
  \delta\overline\phi=V\partial\overline\phi,
\ee
$$
where $h$ is the conformal weight of the generic field $\phi$.

In the path integral formulation [\ANOMALY], the anomaly is generally
 ascribed to a non-invariance of the integration measure and the
 Jacobian factor associated with the anomalous transformation gives
 rise to the anomaly. The Jacobian factor of general conformal fields
 in a conformally flat background, is analyzed in [\YOUR]. According to
 [\YOUR], under an infinitesimal change of the integration variable,
 $\tilde\phi\rightarrow\tilde\phi+\varepsilon(x)\tilde\phi$, a
 logarithm of the Jacobian factor $J$ is given by
$$
  \ln J=\pm{1\over2\pi}\int d^2x\varepsilon(x)
  \left[\left(a-b\over3\right)\partial\partialbar\sigma
  +M^2e^{-(2a+b)\sigma}\right],
\eqn\twoeight
$$
and, for
 $\tilde\phi\rightarrow\tilde\phi+\varepsilon(x)\partial\tilde\phi$,
$$
\eqalign{
  &\ln J
  =\pm{1\over24\pi}\int d^2x\varepsilon(x)
  \biggl[\left(b^2-4ab\right)\partial\sigma\partialbar\partial\sigma
  +\left(2a-3b\right)\partialbar\partial^2\sigma
\cr
  &\qquad\qquad\qquad\qquad\qquad\qquad\qquad\qquad
  -12M^2\left(a+b\right)\partial\sigma e^{-(2a+b)\sigma}\biggr].
\cr
}
\eqn\twonine
$$
In \twoeight~and~\twonine, the double sign~($\pm$) corresponds to the
 statistics of the field~$\phi$. The trace operation that is necessary
 to evaluate the above Jacobian factors is regularized by using an
 exponential type damping factor $e^{-H/M^2}$ with the regulator $H$,
 which is defined by
$$
  H\equiv-\Dslash^\dagger\Dslash
  \equiv-e^{a\sigma}\partial e^{b\sigma}\partialbar e^{a\sigma}.
\eqn\twonines
$$
where $\Dslash$ is the kinetic operator of the field $\tilde\phi$. By
 rewriting the action in~\twofour\ in terms of the integration
 variables in \twofive, we can read off the various values of $a$ and
 $b$ in \twonines\ for the each fields (see Table 1).

To see how our present formulation works, let us first consider the
 ghost number anomaly [\FUJ]. In $N=2$ case, the ghost number current
 is known to be anomaly free, due to a
 cancellation of the background charge [\FRI]. The ghost number
 supercurrent [\GOM] is defined by
$$
\eqalign{
  j_{\rm gh}(Z)&\equiv-BC(Z)
\cr
  &=i\eta c+\thetaminus\left(-\eta\gammaplus+i\betaplus c\right)
   +\thetaplus\left(\eta\gammaminus+i\betaminus c\right)
\cr
  &\quad+\thetaminus\thetaplus\left(-\eta\xi-\betaminus\gammaplus
   -\betaplus\gammaminus-bc\right)
\cr
  &\equiv
  j_{\rm gh}^0(z)+\thetaminus j_{\rm gh}^+(z)
  +\thetaplus j_{\rm gh}^-(z)+\thetaminus\thetaplus
  j_{\rm gh}^{+-}(z).
\cr
}
\eqn\twonineprime
$$

We can immediately see
$$
  \partialbar\VEV{j_{\rm gh}^0(z)}
  =\partialbar\VEV{j_{\rm gh}^+(z)}
  =\partialbar\VEV{j_{\rm gh}^-(z)}=0,
\eqn\twooneone
$$
i.e., these currents are anomaly free. To show this, let us consider
 the
 following infinitesimal change of variables in \twofour:
$$
  b\rightarrow b+i\varepsilon(x)\eta,\quad
  \xi\rightarrow\xi-i\varepsilon(x)c.
\eqn\twoeleven
$$
Note that the partition function itself does not change under a change
 of the integration variables. Therefore the variation of the action
 and the variation of the integration measure should be canceled each
 other, and we have the following identity:
$$
  -{1\over2\pi}\int d^2x\varepsilon(x)
    \partialbar\VEV{j_{\rm gh}^0(z)}
  +\VEV{\ln J}=0,
\ee
$$
where $J$ is a Jacobian factor associated with the change of variables
 \twoeleven. However, for \twoeleven, the Jacobian is trivial, i.e.,
 $J=1$ and
 $\ln J=0$, because the variation of the field is not proportional to
 the field itself.
 Therefore $\partialbar\VEV{j_{\rm gh}^0(z)}$ is
 anomaly free. Similar considerations show other relations in
 \twooneone.

A potential anomalous term in a vacuum expectation value is thus a
 product of the equation of motion and the conjugate field, because
 such a combination is proportional to
 the Jacobian factor of a change of variable whose variation is
 proportional to the field itself. In this sense
 the final combination $\partialbar\VEV{j_{\rm
 gh}^{+-}(z)}$ is potentially dangerous. Let us consider the following
 change of variables:
$$
\eqalign{
  &\delta c=\varepsilon(x)c,\quad\delta b=-\varepsilon(x)b,
\cr
  &\delta\gamma^\pm=\varepsilon(x)\gamma^\pm,\quad
  \delta\beta^\pm=-\varepsilon(x)\beta^\pm,
\cr
  &\delta\xi=\varepsilon(x)\xi,\quad
  \delta\eta=-\varepsilon(x)\eta,
\cr
}
\eqn\twothirteen
$$
(more precisely, we should write down the variation of the tilded
 integration variables, but for~\twothirteen, the variation of the
 tilded
 variables is proportional to the one of the untilded variables). Then
 we have
$$
  -{1\over2\pi}\int d^2x\varepsilon(x)
    \partialbar\VEV{j_{\rm gh}^{+-}(z)}+\VEV{\ln J}=0,
\ee
$$
where $J$ is the Jacobian factor associated with the
 variation~\twothirteen.
 From the master formula \twoeight, we have
$$
  \ln J=-{1\over4\pi}\int d^2x\varepsilon(x)
  (-3+2\times2-1)\partialbar\partial\sigma=0,
\ee
$$
where the contributions from the different sector [($b$,$c$),
 ($\beta^\mp$,$\gamma^\pm$), and ($\eta$,$\xi$) respectively] are
 separately indicated. We can see that the ghost number anomaly
 vanishes due to a cancellation of the background charges. Our
 formulation reproduces the desired answer as is expected.

Let us now turn to the anomaly associated with the  conservation of the
 BRST supercurrent. We define the BRST supercurrent as [\OUR]
$$
\eqalign{
  j_B(Z)&\equiv C\left(T^X+{1\over2}T^{\rm gh}\right)
  +{1\over4}\Diminus\left[C\left(\Diplus C\right)B\right]
  +{1\over4}\Diplus\left[C\left(\Diminus C\right)B\right]
\cr
  &\equiv J_B(Z)+\widehat\jmath_B(Z).
\cr
}
\eqn\twosixteen
$$
To see the structure of the BRST anomaly, we call the first term in the
 first line in \twosixteen\ as $J_B(Z)$ and the total divergence parts
 as $\widehat\jmath_B(Z)$. We should note here that we chose the total
 divergence parts $\widehat\jmath_B(B)$ (which do not affect to the
 BRST charge $Q_B$) by
$$
  j_B(Z)=-\left\{Q_B,j_{\rm gh}(Z)\right\},
\eqn\twoseventeen
$$
to make the BRST current manifestly BRST invariant if $Q_B^2=0$. In
 this sense the above choice is the most symmetric one and this choice
 of $\widehat\jmath_B(Z)$ is crucial for our conclusion.

We also define the components of the BRST supercurrent as
$$
\eqalign{
  &J_B(Z)\equiv J_B^0(z)+\thetaminus J_B^++\thetaplus J_B^-(z)
  +\thetaminus\thetaplus J_B^{+-}(z),
\cr
  &\widehat\jmath_B(Z)\equiv\widehat\jmath_B^0(z)+\thetaminus\widehat
\jmath_B^+
  +\thetaplus\widehat\jmath_B^-(z)
  +\thetaminus\thetaplus\widehat\jmath_B^{+-}(z).
\cr
}
\ee
$$

The explicit form of the component currents of $J_B(Z)$ becomes:
$$
\eqalign{
  &J_B^0(z)={1\over2}c\left[\psiminusmu\psiplusmu
   -i\partial(c\eta)+{1\over2}\gammaplus\betaminus
   -{1\over2}\gammaminus\betaplus\right],
\cr
  &J_B^+(z)=-{1\over2}c\biggl[-i\partial\Ymu\psiplusmu
   +\partial\Xmu\psiplusmu-\partial\left(\gammaplus\eta\right)
   +i\partial\left(c\betaplus\right)-{1\over2}i\gammaplus b
   +{1\over2}\gammaplus\partial\eta
\cr
  &\qquad\qquad\qquad\qquad\qquad\qquad\qquad\qquad\qquad
   \qquad\qquad\qquad+{1\over2}\xi\betaplus
   +{1\over2}i\partial c\betaplus\biggr]
\cr
  &\qquad\qquad-{1\over2}i\gammaplus\left[\psiminusmu\psiplusmu
   -i\partial(c\eta)+{1\over2}\gammaplus\betaminus
   -{1\over2}\gammaminus\betaplus\right],
\cr
  &J_B^-(z)=-{1\over2}c\biggl[-i\partial\Ymu\psiminusmu
   -\partial\Xmu\psiminusmu+\partial\left(\gammaminus\eta\right)
   +i\partial\left(c\betaminus\right)-{1\over2}i\gammaminus b
   -{1\over2}\gammaminus\partial\eta
\cr
  &\qquad\qquad\qquad\qquad\qquad\qquad\qquad\qquad\qquad
   \qquad\qquad\qquad-{1\over2}\xi\betaminus
   +{1\over2}i\partial c\betaminus\biggr]
\cr
  &\qquad\qquad+{1\over2}i\gammaminus\left[\psiminusmu\psiplusmu
   -i\partial(c\eta)+{1\over2}\gammaplus\betaminus
   -{1\over2}\gammaminus\betaplus\right],
\cr
  &J_B^{+-}(z)=\widetilde J_B^{+-}(z)+{3\over4}
   \partial\left(c\gammaplus\betaminus+c\gammaplus\betaminus\right)
   -{1\over2}\partial(c\xi\eta)
\cr
   &\qquad+{i\over2}\xi\left(\psiminusmu\psiplusmu
   +\gammaplus\betaminus-\gammaminus\betaplus\right)
   +{i\over2}\gammaminus\partial\left(\gammaplus\eta\right)
   -{i\over2}\gammaplus\partial\left(\gammaminus\eta\right)
   -{1\over2}\gammaminus\gammaplus b.
\cr
}
\ee
$$
In the last expression, $\widetilde J_B^{+-}(z)$ is defined by
$$
\eqalign{
  \widetilde J_B^{+-}&\equiv
   {1\over2}c\left(\partial\Xmu\partial\Xmu+\partial\Ymu\partial\Ymu
   +\partial\psiminusmu\psiplusmu+\partial\psiplusmu\psiminusmu\right)
\cr
  &\quad+c\left(\partial cb-{1\over2}\gammaminus\partial\betaplus
   -{3\over2}\partial\gammaminus\betaplus
   -{1\over2}\gammaplus\partial\betaminus
   -{3\over2}\partial\gammaplus\betaminus+\partial\xi\eta\right).
\cr
}
\ee
$$
Similarly, the components of the hat supercurrent $\widehat\jmath_B(Z)$
 are given by
$$
\eqalign{
  &\widehat\jmath_B^0(z)=-{i\over2}\gammaminus\gammaplus\eta
   +{1\over4}c\left(\gammaplus\betaminus-\gammaminus\betaplus\right),
\cr
  &\widehat\jmath_B^+(z)=-{1\over4}\partial\left(c\gammaplus\eta\right)
   -{1\over4}c\partial\gammaplus\eta+{i\over2}\gammaplus\xi\eta
   +{1\over4}\gammaplus\partial c\eta
   -{i\over4}\gammaplus\gammaplus\betaminus
\cr
  &\qquad\qquad-{1\over4}c\xi\betaplus
   +{i\over4}c\partial c\betaplus
   -{i\over4}\gammaplus\gammaminus\betaplus+{i\over4}\gammaplus cb,
\cr
  &\widehat\jmath_B^-(z)={1\over4}\partial\left(c\gammaminus\eta\right)
   +{1\over4}c\partial\gammaminus\eta+{i\over2}\gammaminus\xi\eta
   -{1\over4}\gammaminus\partial c\eta
   -{i\over4}\gammaminus\gammaminus\betaplus
\cr
  &\qquad\qquad+{1\over4}c\xi\betaminus
   +{i\over4}c\partial c\betaminus
   -{i\over4}\gammaminus\gammaplus\betaminus+{i\over4}\gammaminus cb,
\cr
  &\widehat\jmath_B^{+-}(z)=-{1\over2}\partial(c\xi\eta)
   +{1\over4}\partial\left(c\gammaplus\betaminus\right)
   +{1\over4}\partial\left(c\gammaminus\betaplus\right).
\cr
}
\eqn\twotwentyone
$$

Let us start the calculation of the BRST anomaly from the first part,
 $\partialbar\VEV{J_B^0(z)}$. Noting the dangerous combination, i.e., a
 product of the equation of motion and the conjugate field, we find
$$
  \partialbar\VEV{J_B^0(z)}
  =\VEV{{1\over2}c\partialbar\left(\psiminusmu\psiplusmu\right)
  +{1\over4}c\partialbar\left(\gammaplus\betaminus%
    -\gammaminus\betaplus\right)}.
\ee
$$
Therefore we use the following variation of the integration variables
$$
  \delta\psiplusmu=-{1\over2}\varepsilon(x)c\psiplusmu,\quad
  \delta\psiminusmu={1\over2}\varepsilon(x)c\psiminusmu,
\eqn\twotwentythree
$$
to get
$$
  -{1\over2\pi}\int d^2x\varepsilon(x)
  \VEV{{1\over2}c\partialbar\left(\psiminusmu\psiplusmu\right)}
  +\VEV{\ln J}=0,
\eqn\twotwentyfour
$$
where $J$ is the Jacobian factor associated with the above
 transformation \twotwentythree. However if we note the fact that, in
 our
 present formulation, the Jacobian factor does not depend on the U(1)
 charge (the superscript $\pm$) but only on the conformal weight, we
 can see the contributions from $\psiplusmu$ and $\psiminusmu$ cancel
 each other, i.e., $\ln J=0$ in~\twotwentyfour. Therefore
$$
  \VEV{{1\over2}c\partialbar\left(\psiminusmu\psiplusmu\right)}=0.
\eqn\twotwentyfive
$$
{}From the same reason, we have
$$
  \VEV{{1\over4}c\partialbar\left(\gammaplus\betaminus\right)}
  =\VEV{{1\over4}c\partialbar\left(\gammaminus\betaplus\right)}.
\eqn\twotwentysix
$$
Combining \twotwentyfive~and~\twotwentysix,
$$
  \partialbar\VEV{J_B^0(z)}=0,
\eqn\twotwentyseven
$$
i.e., $J_B^0(z)$ is anomaly free.

For $\partialbar\VEV{J_B^+(z)}$, since
 $\VEV{-{1\over2}i\gammaplus\partialbar\left(\psiminusmu\psiplusmu
\right)}=0$, we can see,
$$
  \partialbar\VEV{J_B^+(z)}
  =\VEV{{1\over4}i\gammaplus\partialbar(cb)
  -{1\over4}i\partialbar\left(\gammaplus\gammaplus\betaminus\right)
  +{1\over4}i\gammaplus\partialbar\left(\gammaminus\betaplus\right)}.
\ee
$$
 We consider the following variations:
$$
\eqalign{
  &\delta c={1\over4}i\varepsilon(x)\gammaplus c,\quad
  \delta b=-{1\over4}i\varepsilon(x)\gammaplus b,
\cr
  &\delta\gammaplus={1\over4}i\varepsilon(x)\gammaplus\gammaplus,\quad
  \delta\gammaminus=-{1\over4}i\varepsilon(x)\gammaplus\gammaminus,
\cr
  &\delta\betaplus={1\over4}i\varepsilon(x)\gammaplus\betaplus,\quad
  \delta\betaminus=-{1\over4}i\varepsilon(x)\gammaplus\betaminus,
\cr
}
\ee
$$
and obtain the following identity:
$$
  -{1\over2\pi}\int d^2x\varepsilon(x)\partialbar\VEV{J_B^+(z)}
  +\VEV{\ln J}=0.
\ee
$$
{}From the master formula \twoeight, we have
$$
  \ln J={1\over2\pi}\int d^2x\varepsilon(x)
  {i\over8}\gammaplus\partialbar\partial\sigma,
\ee
$$
and then
$$
 \partialbar\VEV{J_B^+(z)}
 ={i\over8}\VEV{\gammaplus\partialbar\partial\sigma}
 ={i\over8}\partialbar\VEV{\gammaplus\partial\sigma}.
\eqn\twothirtytwo
$$
In deriving the last expression, we used a safe equation of motion,
 $\VEV{\partialbar\gammaplus}=0$. The equation of motion alone is
 always
 safe, i.e., the Schwinger--Dyson equation always is valid.

Similarly, for $\partialbar\VEV{J_B^-(z)}$ (by interchanging
 $+\leftrightarrow-$), we have
$$
 \partialbar\VEV{J_B^-(z)}
 ={i\over8}\VEV{\gammaminus\partialbar\partial\sigma}
 ={i\over8}\partialbar\VEV{\gammaminus\partial\sigma}.
\eqn\twothirtythree
$$

To evaluate $\partialbar\VEV{J_B^{+-}(z)}$, let us first consider
 $\partialbar\VEV{\widetilde J_B^{+-}(z)}$. We take the following
 variations:
$$
\eqalign{
  &\delta\sigma=0,
\cr
  &\delta\Xmu=\varepsilon(x)c\partial\Xmu,\quad
   \delta\Ymu=\varepsilon(x)c\partial\Ymu,
\cr
  &\delta\psi^{\pm\mu}
  =\varepsilon(x)\left[c\partial\psi^{\pm\mu}
    +{1\over2}(\partial c)\psi^{\pm\mu}\right],
\cr
  &\delta c=\varepsilon(x)c\partial c,
\cr
  &\delta b=\varepsilon(x)\biggl[
            c\partial b+2(\partial c)b
\cr
  &\qquad\qquad
   +{1\over2}\left(\partial\Xmu\partial\Xmu
     +\partial\Ymu\partial\Ymu+\partial\psiminusmu\psiplusmu
     +\partial\psiplusmu\psiminusmu\right)
\cr
  &\qquad\qquad
   -{1\over2}\gammaminus\partial\betaplus
   -{3\over2}\partial\gammaminus\betaplus
   -{1\over2}\gammaplus\partial\betaminus
   -{3\over2}\partial\gammaplus\betaminus
   +\partial\xi\eta\biggr],
\cr
  &\delta\gamma^\pm=\varepsilon(x)\left[
   c\partial\gamma^\pm-{1\over2}(\partial c)\gamma^\pm\right],
\cr
  &\delta\beta^{\pm}=\varepsilon(x)\left[
   c\partial\beta^\pm+{3\over2}(\partial c)\beta^\pm\right],
\cr
  &\delta\xi=\varepsilon(x)c\partial\xi,
\cr
  &\delta\eta=\varepsilon(x)\left[
   c\partial\eta+(\partial c)\eta\right].
\cr
}
\eqn\twothirtyfour
$$
Then we have the following identity:
$$
\eqalign{
  &-{1\over2\pi}\int d^2x
   \biggl\{\varepsilon(x)\partialbar\VEV{\widetilde J_B^{+-}(z)}
\cr
  &\qquad\qquad\quad-\partial\varepsilon(x)c\bigl[
  {1\over2}\left(\psiplusmu\partialbar\psiminusmu
                 +\psiminusmu\partialbar\psiplusmu\right)
  +b\partialbar c+\eta\partialbar\xi
\cr
  &\qquad\qquad\quad%
  +{3\over2}\betaplus\partialbar\gammaminus
  +{1\over2}\partialbar\betaplus\gammaminus
  +{3\over2}\betaminus\partialbar\gammaplus
  +{1\over2}\partialbar\betaminus\gammaplus\bigr]\biggr\}
\cr
  &\quad+\VEV{\ln J_1}=0,
\cr
}
\eqn\twothirtyfive
$$
where $J_1$ is the Jacobian factor associated the
 variations~\twothirtyfour. The variations \twothirtyfour\ cause the
 following variations of the integration variables:
$$
\eqalign{
  &\delta\widetilde X^\mu=\varepsilon(x)\left[
    c\partial\widetilde X^\mu
    -{1\over2}c(\partial\sigma)\widetilde X^\mu\right],
\cr
  &\delta\widetilde Y^\mu=\varepsilon(x)\left[
    c\partial\widetilde Y^\mu
    -{1\over2}c(\partial\sigma)\widetilde Y^\mu\right],
\cr
  &\delta\widetilde\psi^{\pm\mu}
  =\varepsilon(x)\left[c\partial\widetilde\psi^{\pm\mu}
    -{1\over4}c(\partial\sigma)\widetilde\psi^{\pm\mu}
    +{1\over2}(\partial c)\widetilde\psi^{\pm\mu}\right],
\cr
  &\delta\widetilde c
   =\varepsilon(x)e^{-\sigma}\widetilde c\partial\widetilde c,
\cr
  &\delta\widetilde b=\varepsilon(x)\left[
    c\partial\widetilde b
    +{1\over2}c(\partial\sigma)\widetilde b
    +2(\partial c)\widetilde b\right],
\cr
  &\delta\widetilde\gamma^\pm=\varepsilon(x)\left[
    c\partial\widetilde\gamma^\pm
    -{3\over4}c(\partial\sigma)\widetilde\gamma^\pm
    -{1\over2}(\partial c)\widetilde\gamma^\pm\right],
\cr
  &\delta\widetilde\beta^{\pm}=\varepsilon(x)\left[
    c\partial\widetilde\beta^\pm
    +{1\over4}c(\partial\sigma)\widetilde\beta^{\pm}
    +{3\over2}(\partial c)\widetilde\beta^\pm\right],
\cr
  &\delta\widetilde\xi=\varepsilon(x)\left[
    c\partial\widetilde\xi
    -{1\over2}c(\partial\sigma)\widetilde\xi\right],
\cr
  &\delta\eta=\varepsilon(x)\left[
   c\partial\eta+(\partial c)\eta\right].
\cr
}
\ee
$$
{}From the master formula \twoeight\ and \twonine\ we have
$$
\eqalign{
  &\ln J_1
\cr
  &={1\over2\pi}\int d^2x\varepsilon(x)
  \left[-{d\over3}c\partialbar\partial^2\sigma
         +{d-2\over4}c\partial\sigma\partialbar\partial\sigma
         -{d+6\over12}\partial c\partialbar\partial\sigma
         -dM^2\partial\left(ce^\sigma\right)\right].
\cr
}
\eqn\twothirtyseven
$$
For the remaining part in~\twothirtyfive, we can see
$$
\eqalign{
  &-{1\over2\pi}\int d^2x\left(-\partial\varepsilon(x)\right)
   \biggl\langle
   c\biggl[
   {1\over2}\left(\psiplusmu\partialbar\psiminusmu
                  +\psiminusmu\partialbar\psiplusmu\right)
   +b\partialbar c+\eta\partialbar\xi
\cr
  &\qquad\qquad\qquad\qquad\qquad
               +{3\over2}\betaplus\partialbar\gammaminus
               +{1\over2}\partialbar\betaplus\gammaminus
               +{3\over2}\betaminus\partialbar\gammaplus
               +{1\over2}\partialbar\betaminus\gammaplus
     \biggr]\biggr\rangle
\cr
  &\quad+\VEV{\ln J_2}=0,
\cr
}
\ee
$$
where the Jacobian $J_2$ is associated with
$$
\eqalign{
  &\delta\psi^{\pm\mu}=-{1\over2}\partial\varepsilon(x)c\psi^{\pm\mu},
\cr
  &\delta b=-\partial\varepsilon(x)cb,
\cr
  &\delta\gamma^\pm={1\over2}\partial\varepsilon(x)c\gamma^\pm,\quad
   \delta\beta^\pm=-{3\over2}\partial\varepsilon(x)c\beta^\pm,
\cr
  &\delta\eta=-\partial\varepsilon(x)c\eta.
\cr
}
\ee
$$
By using the master formula,
$$
  \ln J_2={1\over2\pi}\int d^2x\varepsilon(x)
   \left[-{d-12\over12}\partial\left(c\partialbar\partial\sigma\right)
   -dM^2\partial\left(ce^\sigma\right)\right].
\eqn\twoforty
$$
Combining the above results \twothirtyseven~and~\twoforty, we have
$$
  \partialbar\VEV{\widetilde J_B^{+-}(z)}
   ={d-2\over4}\VEV{c\left(\partial\sigma\partialbar\partial\sigma
                      -\partialbar\partial^2\sigma\right)}
    -{3\over2}\VEV{\partial\left(c\partialbar\partial\sigma\right)}.
\ee
$$

Finally we have to calculate:
$$
\eqalign{
  &\partialbar\VEV{J_B^{+-}(z)-\widetilde J_B^{+-}(z)}
\cr
  &=\partialbar\VEV{
   {3\over4}\partial\left(c\gammaminus\betaplus
                          +c\gammaplus\betaminus\right)
    -{1\over2}\partial(c\xi\eta)}
\cr
  &=\partial\VEV{{3\over4}c\partialbar\left(\gammaminus\betaplus
                                     +\gammaplus\betaminus\right)
      -{1\over2}c\partialbar(\xi\eta)}
\cr
  &={5\over4}\VEV{\partial\left(c\partialbar\partial\sigma\right)}.
\cr
}
\ee
$$
(The calculation is similar to the ghost number anomaly.) Collecting
 the above considerations, we finally get
$$
\eqalign{
  \partialbar\VEV{J_B^{+-}(z)}
  &={d-2\over4}\VEV{c\left(\partial\sigma\partialbar\partial\sigma
                           -\partialbar\partial^2\sigma\right)}
   -{1\over4}\VEV{\partial\left(c\partialbar\partial\sigma\right)}
\cr
  &={d-2\over4}\partialbar\VEV{c
             \left({1\over2}\partial\sigma\partial\sigma
                           -\partial^2\sigma\right)}
   -{1\over4}\partialbar\VEV{\partial\left(c\partial\sigma\right)},
\cr
}
\eqn\twofortythree
$$
where, in the final step, we used a safe equation of motion
 $\VEV{\partialbar c}=0$.

For the hat currents in~\twotwentyone, similar calculations show,
$$
\eqalign{
  &\partialbar\VEV{\widehat\jmath_B^0}=0,
\cr
  &\partialbar\VEV{\widehat\jmath_B^+}
    =-{i\over8}\VEV{\gammaplus\partialbar\partial\sigma}
    =-{i\over8}\partialbar\VEV{\gammaplus\partial\sigma},
\cr
  &\partialbar\VEV{\widehat\jmath_B^-}
    =-{i\over8}\VEV{\gammaminus\partialbar\partial\sigma}
    =-{i\over8}\partialbar\VEV{\gammaminus\partial\sigma},
\cr
  &\partialbar\VEV{\widehat\jmath_B^{+-}}
    ={1\over4}\VEV{\partial\left(c\partialbar\partial\sigma\right)}
    ={1\over4}\partialbar\VEV{\partial\left(c\partial\sigma\right)},
\cr
}
\eqn\twofortyfour
$$
where, in the final step, we used safe equations of motion,
 $\VEV{\partialbar\gamma^\pm}=\VEV{\partialbar c}=0$.

One may summarize those identities \twotwentyseven, \twothirtytwo,
 \twothirtythree, and \twofortythree,
 and \twofortyfour\ in supercurrent forms:
$$
\eqalign{
  &\partialbar\VEV{J_B(Z)}
  =\partialbar\biggl\langle
    {i\over8}\thetaminus\gammaplus\partial\sigma
    +{i\over8}\thetaplus\gammaminus\partial\sigma
\cr
    &\qquad\qquad\qquad\qquad
    +\thetaminus\thetaplus\left[
      {d-2\over4}c\left({1\over2}\partial\sigma\partial\sigma
             -\partial^2\sigma\right)
      -{1\over4}\partial\left(c\partial\sigma\right)\right]
      \biggr\rangle,
\cr
  &\partialbar\VEV{\widehat\jmath_B(Z)}
   =\partialbar\VEV{
    -{i\over8}\thetaminus\gammaplus\partial\sigma
    -{i\over8}\thetaplus\gammaminus\partial\sigma
    +{1\over4}\thetaminus\thetaplus
     \partial\left(c\partial\sigma\right)}.
\cr
}
\eqn\twofortyfive
$$

Now, as noted previously, we assumed that our regularization actually
 preserves the global supersymmetry \twotwo. In terms of the
 superfield,
 this assumption requires the right hand sides of~\twofortyfive\ should
 behave
 as covariant supercurrents. Thus here we introduce the super-Liouville
 field by
$$
  \Phi(Z)=\phi(z)+\theta^-\phi^+(z)
          +\theta^+\phi^-(z)+i\theta^-\theta^+\partial\sigma(z),
\ee
$$
where $\phi$ and $\phi^\pm$ are the remaining components of the U(1)
 gauge field and the gravitino respectively, and $\sigma$ is the
 Liouville mode. The covariant combinations which
 reproduce~\twofortyfive\ under a condition $\phi=\phi^\pm=0$ are
$$
\eqalign{
  &\partialbar\VEV{J_B(Z)}
  ={d-2\over4}\partialbar\VEV{C\left({1\over2}D^-\Phi D^+\Phi
                  +i\partial\Phi\right)}
\cr
  &\qquad\qquad\qquad\qquad-{i\over8}\partialbar
   \VEV{\Diplus\left(C\Diminus\Phi\right)
        +\Diminus\left(C\Diplus\Phi\right)},
\cr
  &\partialbar\VEV{\widehat\jmath_B(Z)}
  ={i\over8}\partialbar
   \VEV{\Diplus\left(C\Diminus\Phi\right)
        +\Diminus\left(C\Diplus\Phi\right)}.
\cr
}
\eqn\twofortyseven
$$
The above covariantizations are unique ones. Therefore if we take our
 definition of the BRST supercurrent \twosixteen, we have
$$
  \partialbar\VEV{j_B(Z)}
  ={d-2\over4}\partialbar\VEV{C\left({1\over2}D^-\Phi D^+\Phi
                  +i\partial\Phi\right)}.
\eqn\twofortyeight
$$

Here we should emphasize that the BRST anomaly in \twofortyeight
\ vanishes for
 $d=2$. In the case of $N=0$ and $N=1$ (super-)gravity, on the other
 hand, even if one take a BRST invariant BRST current as in
 \twoseventeen, the
 BRST anomaly contains a additional total divergent piece [\YOU,\YOUR],
 which does not proportional to $d-26$ and $d-10$ respectively. As a
 consequence, the BRST anomaly in $N=0$ and $N=1$ (super-)gravity does
 {\it not\/} vanish even in the critical dimension $d=26$ and $d=10$.
 The origin of the total divergent BRST anomaly is the fact that the
 BRST invariant path integral measure is invariant under the global
 BRST transformation, up to a total divergence [\GRAVITYTWO]. Therefore
 it generates a total divergent anomaly in general under a {\it
 localized\/} BRST transformation, like \twothirtyfour. In this sense,
 the
 absence of a total divergent anomaly in \twofortyeight\ is an
 intrinsic
 feature of the $N=2$ theory and suggests the topological nature of
 $N=2$ theory as a  quantum theory.

We can also summarize the ghost number anomaly in terms of the
 supercurrent:
$$
  \partialbar\VEV{j_{\rm gh}(Z)}=0.
\eqn\twofortynine
$$

The anomalous identities, \twofortyeight~and~\twofortynine\ will play
 an important role when we
 construct the effective covariant supercurrents in $N=2$
 super-Liouville theory.

\chapter{$N=2$ super-Liouville theory and the covariant supercurrents}

In the previous section, we saw that the various anomalies appear
 through the $\sigma$-dependences in the integration measure \twofive.
 Here we try to construct an effective theory which is supposedly
 equivalent with the original $N=2$ supergravity \twofour, by
 incorporating the effect of anomalies. Firstly, following a standard
 procedure to produce the Wess--Zumino term in the string theory
 [\FUJ],
 we repeat an infinitesimal transformation of the integration
 variables. For example, we change $\widetilde X^\mu$ as
$$
  \widetilde X^\mu\longrightarrow
  \left(1+{\sigma\over2}dt\right)\widetilde X^\mu.
\eqn\threeone
$$
By repeating this infinitesimal transformation up to a finite $t$, the
 kinetic operator of $\widetilde X^\mu$ changes to
$$
  e^{-\sigma(1-t)/2}\partial\partialbar e^{-\sigma(1-t)/2},
\ee
$$
thus all the $\sigma$ dependences in \twofour~and~\twofive\ disappear
 at
 $t=1$.

On the other hand, from the master formula \twoeight, the change of
 variable in \threeone\ generates the following Jacobian:
$$
  \ln J(t)_{\widetilde X^\mu}={d\over2\pi}dt\int d^2x\sigma\left[
  -{1\over12}(1-t)\partialbar\partial\sigma+{1\over2}M^2e^{(1-t)\sigma}
  \right].
\ee
$$
Summing over all the contributions from various fields and by
 integrating $\ln J(t)$ from $t=0$ to $1$, we have the so-called
 Liouville action:
$$
  \int_0^1\ln J(t)=-{2-d\over16\pi}
            \int d^2x\partial\sigma\partialbar\sigma.
\eqn\threefour
$$
Note that the ``Liouville term,'' $e^\sigma$, disappears in~\threefour
\ because of the supersymmetry of the original model.

As noted in the previous section, the Liouville action should invariant
 under the global super transformation \twotwo. Therefore, under a
 supersymmetric regularization, the Liouville action \threefour\ should
 have
 the form [\FRAD],
$$
  S_{\rm Liouville}\equiv-{2-d\over16\pi}
            \int d^2x\left[\partial\sigma\partialbar\sigma
                           +\partial\phi\partialbar\phi
                           -\phi^+\partialbar\phi^-
                           -\phi^-\partialbar\phi^+
                           +({\rm c.\ c.})\right].
\eqn\threefive
$$

In this stage, since we have extracted the $\sigma$-dependences in the
 integration measure as the Liouville action \threefive, our partition
function of the matter and the ghost multiplets in a fixed metric
 background has the following form:
$$
\eqalign{
  \int\,d\mu\,\exp\biggl\{-{1\over2\pi}\int\biggl[\,
  {1\over2}&\left(
  -\partial\Xmu\partialbar\Xmu
  -\partial\Ymu\partialbar\Ymu
  +\psiplusmu\partialbar\psiminusmu+\psiminusmu\partialbar\psiplusmu
  \right)
\cr
  &\qquad+b\partialbar c+\betaplus\partialbar\gammaminus
   +\betaminus\partialbar\gammaplus+\eta\partialbar\xi
   +({\rm c.\ c.})\,\biggr]\biggr\},
\cr
}
\eqn\threesix
$$
where $d\mu$ is a ``naive'' integration measure:
$$
\eqalign{
  d\mu&=
  \Di c
  \Di\overline c
  \Di b
  \Di\overline b
  \Di\gammaplus
  \Di\overline\gammaplus
  \Di\gammaminus
  \Di\overline\gammaminus
  \Di\betaplus
  \Di\overline\betaplus
  \Di\betaminus
  \Di\overline\betaminus
\cr
  &\quad\times
  \Di\xi
  \Di\overline\xi
  \Di\eta
  \Di\overline\eta
  \Di\psiplusmu
  \Di\overline\psiplusmu
  \Di\psiminusmu
  \Di\overline\psiminusmu
  \Di\Xmu
  \Di\Ymu.
\cr
}
\eqn\threeseven
$$
{}From \threesix~and~\threeseven, we have the following correlation
 functions of $X^\mu(Z)$, $C(Z)$, and $B(Z)$:
$$
\eqalign{
 &\VEV{X^\mu(Z_a)X^\nu(Z_b)}=\eta^{\mu\nu}\ln Z_{ab},
\cr
 &\VEV{C(Z_a)B(Z_b)}=\VEV{B(Z_a)C(Z_b)}=
 {{\theta_{ab}^-\theta_{ab}^+}\over{Z_{ab}}},
\cr
}
\eqn\threeeight
$$
where $Z_{ab}$ and $\theta^\pm_{ab}$ are defined by
$$
\eqalign{
 &Z_{ab}
  =z_a-z_b-\left(\theta_a^+\theta_b^-+\theta_a^-\theta_b^+\right),
\cr
 &\theta^\pm_{ab}=\theta^\pm_a-\theta^\pm_b.
\cr
}
\ee
$$

Our next question is the following: What is the correct expression of
 the ghost number supercurrent and the BRST supercurrent in the
 partition function \threesix? Note that, in the partition
 function~\threesix,
 we do not have any anomalies and we can always use the naive equations
 of motion. From the expressions of the BRST anomaly~\twofortyseven
\ and the
 ghost number anomaly~\twofortyeight, we may take
$$
\eqalign{
  &j_B(Z)\equiv C\left(T^X+{1\over2}T^{\rm gh}\right)
   +{d-2\over4}C
    \left({1\over2}\Diminus\Phi\Diplus\Phi+i\partial\Phi\right)
\cr
  &\qquad\qquad
  +{1\over4}\Diminus\left[C\left(\Diplus C\right)B\right]
  +{1\over4}\Diplus\left[C\left(\Diminus C\right)B\right],
\cr
  &j_{\rm gh}(Z)\equiv-BC,
\cr
}
\eqn\threeten
$$
as the effective covariant supercurrents in the partition
 function~\threesix. In \threeten, we determined the $\Phi$-dependences
 as to
 reproduce \twofortyeight~and~\twofortynine\ under uses of the naive
 equations of
 motion of the matter and the ghost fields. This prescription was also
 applied to $N=0$ and $N=1$ (super-)gravity [\YOU,\YOUR]. We emphasize
 that
 we obtained the anomalous identities \twofortyeight~and~\twofortynine
\ in the BRST
 invariant path integral framework [\YASUDA,\GRAVITYTWO], thus those
 expressions
 should reflect the (super-)coordinate covariance in the quantum
 theory.

In order to have a complete description of the $N=2$ quantum
supergravity, we should quantize the Liouville supermultiplet. We
 define
 the partition function of the Liouville supermultiplet as
$$
\eqalign{
  &\int\Di\left(e^{\sigma/2}\right)
      \Di\left(e^{\sigma/4}\phi^+\right)
      \Di\left(e^{\sigma/4}\overline\phi^+\right)
      \Di\left(e^{\sigma/4}\phi^-\right)
      \Di\left(e^{\sigma/4}\overline\phi^-\right)
      \Di\left(e^{\sigma/2}\phi\right)
\cr
  &\times\exp\left\{
   -{2-d\over16\pi}\int d^2x\left[\partial\sigma\partialbar\sigma
                           +\partial\phi\partialbar\phi
                           -\phi^+\partialbar\phi^-
                           -\phi^-\partialbar\phi^+
                           +({c.\ c.})\right]
   \right\},
\cr
}
\eqn\threeeleven
$$
where we have chosen the weight factors ($\exp\sigma$) following the
 prescription in [\YASUDA,\GRAVITYTWO]. The full partition
 function is given by a
 product of \threesix~and~\threeeleven.

If we apply the same procedure of the derivation of \threefour\ also to
 the
 gravitinos $\phi^\pm$ and the gauge field $\phi$, the partition
 function \threeeleven\ changes to
$$
\eqalign{
  &\int\Di\left(e^{\sigma/2}\right)
      \Di\phi^+\Di\overline\phi^+\Di\phi^-\Di\overline\phi^-\Di\phi
\cr
  &\times\exp\left[
   -\left({2-d\over16\pi}-{1\over24\pi}\right)
    \int d^2x\partial\sigma\partialbar\sigma
   -{1\over4\pi}M^2\int d^2xe^\sigma
   +\cdots
   \right],
\cr
}
\eqn\threetwelve
$$
where we only indicated the $\sigma$-dependence of the action. The
 integration of the Liouville field in \threetwelve\ is, on the other
 hand,
 highly non-linear because the integration variable is $e^{\sigma/2}$,
 not simply $\sigma$. To avoid this problem, here we apply the
 background field method [\YOU,\YOUR] and include the one-loop
 renormalization effect arising from the non-trivial measure
 $e^{\sigma/2}$.

To do this, we set $e^{\sigma/2}\equiv e^{\sigma_0/2}+\varphi$ and
 expand the Liouville action in~\threetwelve\ with respect to $\varphi$
 up to
 the second order. If we assume the $M^2e^\sigma$ term in \threetwelve
\ is
 canceled by a suitable counter term, the resulting action for the
 quantum fluctuation $\varphi$ has the same form of the action of
 $\widetilde X^\mu(z)$ with replacement $\sigma\rightarrow\sigma_0$.
 Thus, up to the one-loop order, we have additional contribution from
 the Liouville part itself,
$$
  {1\over16\pi}\int d^2x
   \partial\sigma_0\partialbar\sigma_0
  .
\ee
$$
We regard this factor as the one-loop finite renormalization effect.
 Adding this effect to the original contribution from \threetwelve, and
 after
 a covariantization, we finally have
$$
\eqalign{
  &\int\Di\sigma
      \Di\phi^+\Di\overline\phi^+\Di\phi^-\Di\overline\phi^-\Di\phi
\cr
  &\times\exp\left\{
   -{1-d\over16\pi}\int d^2x\left[\partial\sigma\partialbar\sigma
                           +\partial\phi\partialbar\phi
                           -\phi^+\partialbar\phi^-
                           -\phi^-\partialbar\phi^+
                           +({\rm c.\ c.})\right]
   \right\},
\cr
}
\eqn\threeforteen
$$
where we have rewritten $\sigma_0$ as $\sigma$ and taken a naive
 integration
 measure $\sigma$ for the Liouville field, since we already include the
 (one-loop) quantum effect of $e^{\sigma/2}$. We should note here the
 coefficient in \threeeleven, $2-d$, changes to $1-d$ in \threeforteen.
 Since the
 action in \threeforteen\ has the same form as the matter
 supermultiplet
 $X^\mu(Z)$, the correlation function of the Liouville superfield
 $\Phi$ is given by
$$
  \VEV{\Phi\left(Z_a\right)\Phi\left(Z_b\right)}
  ={4\over d-1}\ln Z_{ab}.
\eqn\threefifteen
$$

As the effective covariant supercurrents in the partition function
 \threeforteen, we may take \threeten\ with a replacement
 $2-d\rightarrow 1-d$,
 i.e.,
$$
\eqalign{
  &j_B(Z)\equiv C\left(T^X+{1\over2}T^{\rm gh}\right)
   +{d-1\over4}C
    \left({1\over2}\Diminus\Phi\Diplus\Phi+i\partial\Phi\right)
\cr
  &\qquad\qquad
  +{1\over4}\Diminus\left[C\left(\Diplus C\right)B\right]
  +{1\over4}\Diplus\left[C\left(\Diminus C\right)B\right],
\cr
  &j_{\rm gh}(Z)\equiv-BC.
}
\eqn\threesixteen
$$
We comment on the differences of \threesixteen\ from the analogous
 construction
 for the $N=0$ and $N=1$ (super-)Liouville cases [\YOU,\YOUR]. The
 differences are i)~no appearance of a correction term due to the
 Liouville field in the ghost number supercurrent $j_{\rm gh}(Z)$ in
 \threesixteen\ and, ii)~no appearance of a divergence correction term
 in the
 expression of BRST supercurrent $j_B(Z)$ in \threesixteen. The origins
 of
 these facts are respectively, i) a vanishing of the ghost number
 anomaly in $N=2$ theory \twofortynine, ii)~no appearance of the BRST
 anomaly which is not proportional to $d-2$ in \twofortyeight. In fact,
 as is
 discussed in
 the following section, these two facts might be related each other.

We regard the whole set of the partition function
 \threesix~and~\threeforteen, and
 the effective supercurrents in \threesixteen\ as the effective theory
 of the
 two-dimensional $N=2$ supergravity, i.e., $N=2$ super-Liouville
 theory. The advantage of this effective theory is that we can use
 propagators in a flat space-time, like \threeeight. Although the
 replacement
 $2-d\rightarrow1-d$ in the current operator construction in
 \threesixteen\ is
 {\it ad hoc}, we will check the covariance of those supercurrents by
 using the operator product expansion (OPE) in the next section. This
 shift of the parameter, $2-d\rightarrow1-d$ also appeared as the
 ansatz in [\DIS].

\chapter{BRST supercurrent algebra and the topological conformal
 algebra}

In this section, we show that our effective supercurrents in
 \threesixteen\ forms a topological conformal algebra [\EGU,\NOJ],
 which appears in
 two-dimensional topological (conformal) field theories [\WITT,\WIT].
 This
 observation was reported in our previous communication [\OUR].

Firstly we change the normalization of the Liouville superfield as
$$
  \Phi(Z)\longrightarrow{2\over\sqrt{d-1}}\Phi(Z).
\ee
$$
Thus the correlation function in \threefifteen\ changes to
$$
  \VEV{\Phi\left(Z_a\right)\Phi\left(Z_b\right)}
  =\ln Z_{ab},
\ee
$$
and the effective BRST supercurrent changes to
$$
\eqalign{
 j_B(Z)&=C(Z)\left(T^X+T^{\rm Liouville}+{1\over2}T^{\rm gh}\right)
\cr
 &\quad+{1\over4}D^-\left[C\left(D^+C\right)B\right]
       +{1\over4}D^+\left[C\left(D^-C\right)B\right].
\cr
}
\ee
$$
In the above expression, we defined the Liouville energy-momentum
 tensor:
$$
 T^{\rm Liouville}={1\over2}D^-\Phi D^+\Phi+\kappa\partial\Phi,
\ee
$$
where $\kappa$ satisfies
$$
 \kappa^2={{1-D}\over4}.
\ee
$$

The BRST charge in the $N=2$ super-Liouville theory thus is given by
$$
 Q_B=\int DZ\,C\left(T^X+T^{\rm Liouville}+{1\over2}T^{\rm gh}\right),
\ee
$$
and, as we will see, it satisfies $Q_B^2=0$ for any $d$. Furthermore we
 will also see
$$
 T(Z)=\left\{Q_B,B(Z)\right\},
\eqn\fourseven
$$
where the total energy momentum tensor $T$ is defined by
$$
 T=T^X+T^{\rm gh}+T^{\rm Liouville}.
\ee
$$

At this point we examine the BRST supercurrent algebra in $N=2$
 super-Liouville theory. We change the notation as
$$
\eqalign{
 &T(Z)\equiv T(Z),
\cr
 &G(Z)\equiv j_B(Z),
\cr
 &\overline G(Z)\equiv B(Z),
\cr
 &J(Z)\equiv j_{\rm ghost}(Z).
\cr
}
\eqn\fournine
$$

For the superconformal properties, the relevant operator product
 expansion is (for any $d$),
$$
\eqalign{
 T(Z_a)\Psi(Z_b)&\sim
 h\,{{\theta_{ab}^-\theta_{ab}^+}\over{Z_{ab}^2}}\,\Psi(Z_b)
 +{1\over{2Z_{ab}}}\left(\theta_{ab}^-D_b^+-\theta_{ab}^+D_b^-\right)
 \Psi(Z_b)
\cr
 &\quad+{{\theta_{ab}^-\theta_{ab}^+}\over{Z_{ab}}}
 \,\partial_{z_b}\Psi(Z_b),
\cr
}
\eqn\fourten
$$
where $\Psi=T$, $G$, $\overline G$, and $J$ with $h=1$, $0$, $1$, and
 $0$ respectively. This expression implies those operators are
 primary fields with the U(1) charge $0$ and the superconformal
 weight
 $h$. Especially the case $\Psi=T$ implies a vanishing of the total
 central charge for any $d$. Moreover the BRST supercurrent $j_B(Z)$
 and the ghost number supercurrent $j_{\rm gh}(Z)$ in \threesixteen
\ are
 primary fields. In this sense, the supercurrents \threesixteen\ in
 this
 effective theory are {\it covariant\/} in the quantum
 level and this desired feature suggests our construction in
 \threesixteen\ is
 reliable.

For another relations between various operators, we have (also for any
 $d$),
$$
\eqalign{
 &G(Z_a)\overline G(Z_b)\sim
 {1\over{2Z_{ab}}}\left(\theta_{ab}^-D_b^+-\theta_{ab}^+D_b^-\right)
 J(Z_b)
 +{{\theta_{ab}^-\theta_{ab}^+}\over{Z_{ab}}}
 \,T(Z_b),
\cr
 &G(Z_a)G(Z_b)\sim0,
\cr
 &\overline G(Z_a)\overline G(Z_b)\sim0,
\cr
 &J(Z_a)J(Z_b)\sim0,
\cr
 &J(Z_a)G(Z_b)\sim
  {{\theta_{ab}^-\theta_{ab}^+}\over{Z_{ab}}}\,G(Z_b),
\cr
 &J(Z_a)\overline G(Z_b)\sim
  -{{\theta_{ab}^-\theta_{ab}^+}\over{Z_{ab}}}\,\overline G(Z_b).
\cr
}
\eqn\foureleven
$$
Surprisingly, in the above operator algebra, no quantum anomalous term
 appears and it coincides with the classically expected form. In the
 case of $N=0$~and~$N=1$ cases [\YOUR,\FUJI], on the other hand, the
 quantum
 anomalous terms vanish only at $d=-2$~and~$d=\pm\infty$ respectively.
 The algebra in \fourten~and~\foureleven\ form a kind of the
 topological
 conformal algebra or the twisted $N=4$ superconformal algebra
 [\EGU,\NOJ].
 Therefore, in the $N=2$ super-Liouville theory, the super-coordinate
 BRST supercurrent algebra gives a representation of a $N=2$ superfield
 extension of the topological conformal algebra for any
 $d$\rlap.\footnote{\star}{By comparing the conformal weight and the
 statistics of the each component fields, we can see that our algebra
 in \fourten~and~\foureleven\ are not the same as the twisted $N=4$
 superconformal
 algebras analyzed by Nojiri [\NOJ].} Our observation thus suggests the
 topological
 nature of the $N=2$ super-Liouville theory (or the $N=2$ fermionic
 string theory) as the quantum theory.

We note that the first relation in~\foureleven\ implies \fourseven, and
 the second
 relation in~\foureleven\ implies the BRST invariance of the BRST
 supercurrent, therefore the BRST charge is nilpotent.

The anomaly-free property of the operator algebra in
 \fourten~and~\foureleven\ might be understood from the absence of the
 ghost number anomaly in $N=2$ theory: In our construction, which is
 analogous to the one in [\YOU,\YOUR], the correction of the ghost
 number current due to the Liouville mode is determined from the ghost
 number anomaly. In $N=2$ case, since we have no ghost number anomaly
 in
 \twofortynine, the form of the effective ghost number supercurrent
 in~\threesixteen\ has no correction due to the Liouville mode. From
 experiences on the $N=0$ and $N=1$ (super-)Liouville theories
 [\YOU,\YOUR,\FUJI], we know the following relation for the {\it
 effective\/} currents:
$$
  j_B=-\left\{Q_B,j_{\rm gh}\right\},
\eqn\fourtwelve
$$
and we can see that the $(d-{\rm critical\ dimensions}+1)$
 non-proportional correction of the BRST (super-)current in the left
 hand side is generated from the Liouville correction of the ghost
 number (super-)current in the right hand side. If we expect the
 general validity of the relation \fourtwelve\ in our construction,
 the BRST
 supercurrent in the $N=2$ super-Liouville theory will not have $d-1$
 non-proportional correction. Actually, we can check \fourtwelve\ from
 the explicit form \threesixteen. In the BRST current algebra like
 \foureleven\ in the $N=0$ and $N=1$ case [\YOU,\YOUR,\FUJI], we can
 also observe that the anomalous terms in the algebra arise from the
 above mentioned non-trivial correction of the BRST and ghost number
 (super-)currents. In the case of $N=2$, therefore, we may expect the
 anomaly free property of the operator algebra.

It is also useful to see how the critical string can be considered
as a subcritical string in dimension 1 plus the Liouville superfield,
 in fact in this situation, $\kappa=0$ and all the operators of the
effective theory coincide with the ones of the critical string, since
 in this case there is no restriction for the possible values of $d$.

\chapter{Conclusion}

We computed the BRST and the ghost number anomalies in the $N=2$
 supergravity in the superconformal gauge, working in a path integral
 in terms of the component fields. The dependences of the Liouville
 mode in the anomalies were directly calculated while the dependences
 of the Liouville superpartners were determined by using the global
 supersymmetry \twotwo. The final results were written in terms of
 superfields.

The effective $N=2$ super-Liouville theory was constructed at one loop
 level and there is a finite renormalization of the coupling constant,
 i.e., from $2-d$ to $1-d$. The algebra of the operators in \fournine
\ gives rise to an $N=2$ superfield extension of the topological
 conformal algebra for any value of dimensions $d$ without any
 anomalous terms. The crucial points for this property are the
 vanishing of the ghost number anomaly and the definition of the BRST
 supercurrent. Our observation shows an appearing of a quite
 simplification in the $N=2$ case and also suggests a topological
 nature of the $N=2$ super-Liouville theory.

The $N=2$ critical string can be considered as an $N=2$ subcritical
 string in dimension $1$ plus the Liouville superfield. All the above
 features distinguish $N=2$ string from the $N=0$ and $N=1$ strings.
 The physical relevance of the topological algebra is under study.

\ack{H. S. would like to thank the members of Department d'Estruc\-tura
 i
 Constituents de la Mat\`eria, Universitat de Barcelona, where the most
 part of the present work was done, for their kind hospitality.}

\appendix{}

Let us recall some basic facts of $N=2$ string in the superfield
 formalism\rlap,\footnote{\star}{We follow the notation of [\GOM].} The
 $N=2$ superspace is described in terms of the bosonic $(z,\overline
 z)$ and the fermionic $(\theta^\pm,\overline\theta^\pm)$ coordinates.
 We define covariant derivatives by
$$
 D^\pm={\partial\over{\partial\theta^\mp}}+\theta^\pm\partial,\quad
 \overline D^\pm={\partial\over{\partial\overline\theta^\mp}}
  +\overline\theta^\pm\overline\partial.
\ee
$$
The action can be written in terms of two superfields
 $S^\mu(z,\overline
 z,\theta^+,\overline\theta^+,\theta^-,\overline\theta^-)$ and
 \break
$S^{\mu\ast}(z,\overline
 z,\theta^+,\overline\theta^+,\theta^-,\overline\theta^-)$ satisfying
 two constraints ($\mu$ runs over $1$ to $d$):
$$
 D^-S^\mu=\overline D^-S^\mu=0,
\ee
$$
and
$$
 D^+S^{\mu\ast}=\overline D^+S^{\mu\ast}=0.
\ee
$$
The action is given by [\ADE]
$$
 A=\int dzd\overline z\int
 d\theta^+d\overline\theta^+d\theta^-d\overline\theta^-
  S^{\mu\ast} S^\mu.
\ee
$$
The solution of the equations of motion
$$
 D^+\overline D^+S^\mu=0,\quad\overline D^-D^-S^{\mu\ast}=0,
\ee
$$
can be written as
$$
 S^\mu=S_1^\mu+S_2^\mu,
\ee
$$
where
$$
\eqalign{
 &D^-S_1^\mu=\overline D^-S_1^\mu=\overline D^+S_1^\mu=0,
\cr
 &D^-S_2^\mu=\overline D^-S_2^\mu=D^+S_2^\mu=0.
\cr
}
\ee
$$
A real superfield $X^\mu$ is constructed via
$$
 X^\mu\left(z,\theta^+,\theta^-\right)
  =S_1^\mu\left(z+\theta^-\theta^+,\theta^-\right)
  +S_1^{\mu\ast}\left(z+\theta^+\theta^-,\theta^+\right).
\ee
$$
The components of $X^\mu(Z)$ are
$$
 X^\mu(Z)=X^\mu(z)+\theta^-\psi^{+\mu}(z)+\theta^+\psi^{-\mu}(z)
  +i\theta^-\theta^+\partial Y^\mu(z),
\ee
$$
where $X^\mu(z)$ and $Y^\mu(z)$ are free bosonic fields and
 $\psi^{\pm\mu}(z)$ are free fermions.

 The contribution to the energy momentum tensor from $X^\mu$ is
$$
T^X(Z)={1\over2}D^-X^\mu D^+X^\mu(Z).
\ee
$$

The $N=2$ string action is invariant under several local gauge
 transformations. We are working in the superconformal gauge. The gauge
 fixing generates a Faddeev--Popov determinant expressible as a
 superfield action using $N=2$ superfield ghost $C$ and antighost $B$:
$$
\eqalign{
 &C\equiv c+i\theta^+\gamma^--i\theta^-\gamma^
  ++i\theta^-\theta^+\xi,
\cr
 &B\equiv-i\eta-i\theta^+\beta^--i\theta^-\beta^+
 +\theta^-\theta^+b.
\cr
}
\ee
$$
The ghosts $c$ and $b$ are for the $\tau$-$\sigma$ general coordinate
 invariances, $\gamma^\pm$ and $\beta^\pm$ are the super ghosts for
 the two local supersymmetry transformations and $\xi$ and $\eta$ are
 the ghosts associated with the local U(1) symmetry. Their
 Lagrangians
 are the first order systems with background charge $Q$ [\FRI] and
 statistics $\epsilon$ of $(Q,\epsilon)=(-3,+)$, $(2,-)$ and $(-1,+)$
 respectively. Notice that the total background ghost charge vanishes.
 The ghost action in terms of superfield is given by
$$
 A_{\rm gh}={1\over\pi}\int d^2zd\theta^+d\theta^-B\overline\partial\,C
 +({\rm c.\ c.}).
\ee
$$
The ghost energy momentum tensor becomes
$$
 T^{\rm gh}(Z)=\partial(CB)(Z)-{1\over2}D^+CD^-B(Z)
 -{1\over2}D^-CD^+B(Z).
\ee
$$

\refout
\vfill\eject
%%%%%%%%%%%%%%%%%%%%%%%%%%%%%%%%%%%%%%%%%%%%%%%%%%%%%%%%%%%%%%%%%%%%
%          table 1
%%%%%%%%%%%%%%%%%%%%%%%%%%%%%%%%%%%%%%%%%%%%%%%%%%%%%%%%%%%%%%%%%%%%
\bigskip
\hskip4em
\vbox{\offinterlineskip
\halign{\hbox{\vrule height15pt depth 10pt width0pt}
        \quad\hfil#\hfil\quad\vrule&\quad\hfil#\hfil\quad
       &\vrule\qquad\hfil#\hfil\qquad &\vrule\qquad\hfil#\hfil\qquad\cr
 &statistics &$a$ &$b$ \cr
\noalign{\hrule}
$\widetilde c$ &$-$ &$-1$ &$1$\cr
$\widetilde b$ &$-$ &$\textstyle{1\over2}$ &$-2$\cr
$\widetilde\gamma^\pm$ &$+$ &$\textstyle{-{3\over4}}$
 &$\textstyle{1\over2}$\cr
$\widetilde\beta^\pm$ &$+$ &$\textstyle{1\over4}$
                               &$\textstyle{-{3\over2}}$\cr
$\widetilde\xi$ &$-$ &$\textstyle{-{1\over2}}$ &$0$\cr
$\eta$ &$-$ &$0$ &$-1$\cr
$\widetilde\psi^\mu$ &$-$ &$\textstyle{-{1\over4}}$
                               &$\textstyle{-{1\over2}}$\cr
$\widetilde X^\mu,\widetilde Y^\mu$ &$+$ &$\textstyle{-{1\over2}}$
                               &$0$\cr
}}
\bigskip
\bigskip
\centerline{Table 1}
\vfill
%%%%%%%%%%%%%%%%%%%%%%%%%%%%%%%%%%%%%%%%%%%%%%%%%%%%%%%%%%%%%%%%%%%%
\bye